# SCIENTIFIC REPORTS

**OPEN**

# Metal-coated microsphere monolayers as surface plasmon resonance sensors operating in both transmission and reflection modes

Cosmin Farcau [1,2]

Metal-coated microsphere monolayers (MCM) are a class of plasmonic crystals consisting of noble metal films over arrays of self-assembled colloidal microspheres. Despite their ease of fabrication and tunable plasmonic response, their optical sensing potential has been scarcely explored. Here, silver coated polystyrene sphere monolayers are proposed as surface plasmon resonance sensors capable of functioning in both transmission (*T*) and reflection (*R*) readout modes. An original and key point is the use of ~200 nm colloids, smaller than in MCM studied before. It allowed us to reveal a previously unobserved, additional/secondary Enhanced Optical Transmission band, which can be exploited in sensing, with higher sensitivity than the better-known main transmission band. The reflection configuration however, is almost an order of magnitude more efficient for sensing than the transmission one. We also evidenced a strong impact of the adsorbate location on the metal surface on the sensing efficiency. Electric field distribution analysis is performed to explain these results. Proof-of-concept experiments on the detection of 11-MUA molecular monolayers, performed in both readout modes, confirm the behaviors observed through FDTD simulations. Results in this paper can serve as guidelines for designing optimized sensors based on metal-coated colloidal monolayers, and more generally for plasmonic sensors based on metal nanostructured films.

Surface plasmon resonance (SPR) - based sensing has emerged as a versatile tool for label-free characterization and quantification of bio-molecular interactions[1]. The sensing mechanism relies on the sensitivity of surface plasmons (collective oscillations of conduction electrons) on noble metal surfaces to media in their nanoscopic vicinity. Commercially available systems, such as Biacore[2], exploit surface plasmons on planar thin gold films, excited through a prism in Kretschmann configuration; the angle of minimum reflectivity is monitored for detecting the sensing events. The study of interactions involving many kind of molecules, proteins, nucleic acids, and even viruses or whole living cells is thus possible, with applications in domains like medicine or ecology[3,4]. In recent years, following the bloom of the field of plasmonics, a wide palette of sensors based on nanostructured metal surfaces have been demonstrated[5,6]. With a morphology contrasting the classical flat-surface plasmonic sensors, these offer several advantages, such as: the sensing area can be localized with higher precision; the optical response (e.g. transmitted or reflected spectrum) can be tuned, so the wavelength of detection can be adjusted; the enhanced electromagnetic fields associated to surface plasmons on nanostructures can be exploited to develop multi-modal detection schemes (e.g. combined SPR and Surface Enhanced Raman Scattering)[7]; the shape and size of the sensing chips and the equipment can be diversified according to specific application needs. Current research challenges regarding nanostructured plasmonic bio-sensors include fundamental understanding of the shape and size dependent optical/plasmonic properties in nanostructures, designing and fabricating optimized nanostructures, improving sensitivity and detection limits[8]. Many kinds of nanostructures have been explored for plasmonic sensing, including colloidal nanoparticles in aqueous suspensions[3], thin films fabricated by electron beam lithography or colloidal lithography[9], nanoimprint lithography[10], and many others[5,6]. Note also that up to now, studies report on plasmonic nanostructures working either in transmission or reflection mode, as imposed by the particular optical properties of the sensing elements[11].

[1]National Institute for Research and Development of Isotopic and Molecular Technologies, 67-103 Donat Str., 400293, Cluj-Napoca, Romania. [2]Institute for Interdisciplinary Research in Bio-Nano-Sciences, Babes-Bolyai University, 42 T. Laurian, 400271, Cluj-Napoca, Romania. Correspondence and requests for materials should be addressed to C.F. (email: cfarcau@itim-cj.ro)

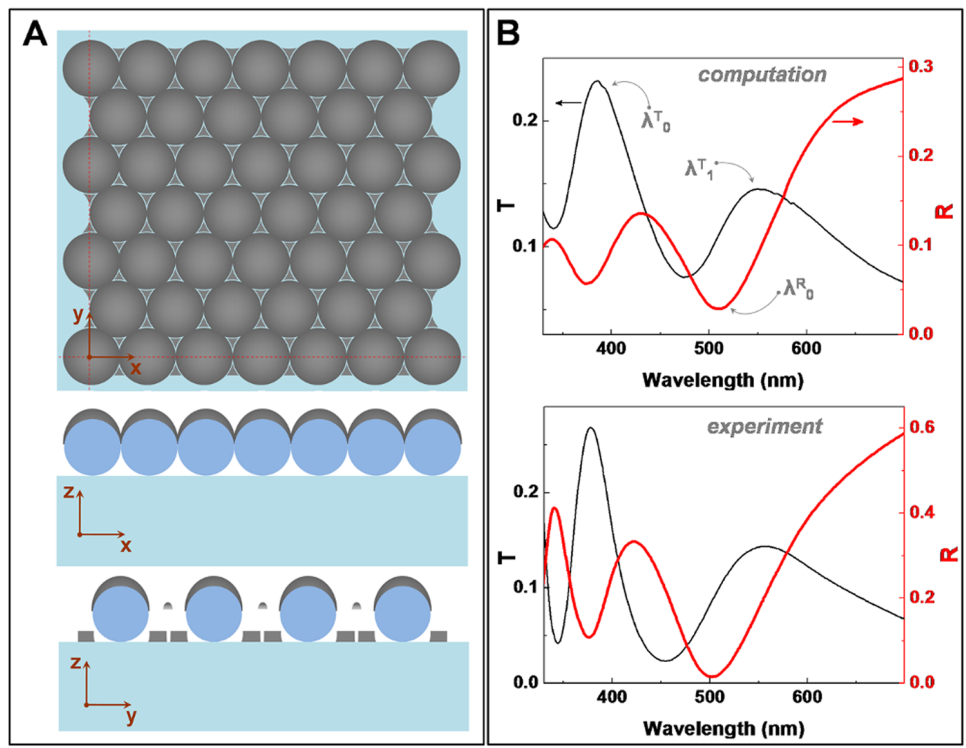

**Figure 1.** (**A**) The simulated structure model: *XY* plane view on top, *XZ* and *YZ* views below, along the planes marked by dashed lines; (**B**) (top) Computed transmittance *T* and reflectance *R* spectra; and (bottom) experimental *T* and *R* spectra.

One particular class of plasmonic crystals that are easy to fabricate and posses high application potential consists of thin metal films (20–200 nm) deposited over an array of dielectric colloidal microspheres (polymer, silica), named hereby metal-coated microsphere monolayers (MCM), and also known as films over nanospheres in the literature[12]. One of the outstanding qualities of MCM is that their optical response can be easily tuned by the sphere diameter and metal film thickness. Another unique feature is that MCM exhibits a plasmon-enhanced transmission band[13–18], similar to the well-known Extraordinary Optical Transmission (EOT)[19]. Furthermore, MCM have been implemented in sensing schemes based on Surface Enhanced Raman Scattering[12] or for amplifying radiative rates as in Metal Enhanced Fluorescence[20]. Finally, although it was suggested that the EOT band on gold-coated microsphere monolayers could be used for sensing in transmission configuration[21], the potential of MCM as SPR sensors has not been further explored.

Here, we employ Finite-Difference Time-Domain computations on realistic three-dimensional structures, to study the sensitivity of MCM as SPR sensors. The optical response of MCM made of small polystyrene spheres (210 nm) coated by silver films is simulated with and without an analyte coating layer (e.g. adsorbate molecules). Due to the small lattice parameter (sphere size), optical spectra rich in spectral features are obtained and analyzed thoroughly for the first time. We question whether both transmittance and reflectance spectra are appropriate for sensing experiments with these MCM structures. Electromagnetic field analysis is then performed to inquire about the distribution of the enhanced plasmonic fields and their impact on the sensor's performance. Further, we show how the precise nanoscopic location of adsorbates on the micro-structure surface influences their detection. The findings of the theoretical analyses are validated by proof-of-concept experiments performed on Ag-coated polystyrene spheres on which 11-MUA molecular monolayer was adsorbed as a model analyte.

## Results and Discussion

**Simulated structure and optical spectral features.** The investigated three dimensional model structure is presented in Fig. 1A, both as a top view and cross-sections along the main vertical planes (indicated by dashed lines). The structure can be viewed as an array of interconnected metallic half-shells superimposed on the array of dielectric spheres. On Fig. 2B the computed (top panel) transmittance and reflectance spectra are presented and compared with typical experimental spectra (bottom panel).

A very good agreement between experiment and theory was obtained, which demonstrates the robustness and reliability of the simulated structure and the used FDTD method, and validates our procedures before proceeding to other, more complex analyses. Note that a very good matching was possible because an extended, finite array was modeled, which contrasts most previous simulations in which a single lattice cell and periodic boundary conditions were used. A minor difference between simulated and experimental spectra is the slightly broader peaks in the simulation, which can be an effect induced by the finite-size lattice. The higher reflectance value in the experiments could be due to cummulative effect of the zero-order correction in the simulation and the reference mirror used in the experiments not providing an absolute reflectance value. Anyway, compared to previous results on similar structures[14,22], but employing periodic boundary conditions, the matching between simulation and experiments is better in our case. Also note that, because of the chosen metal film thickness, the optical spectra present multiple features (minima, maxima) in both transmittance and reflectance spectra, useful for implementing multiple sensing protocols.

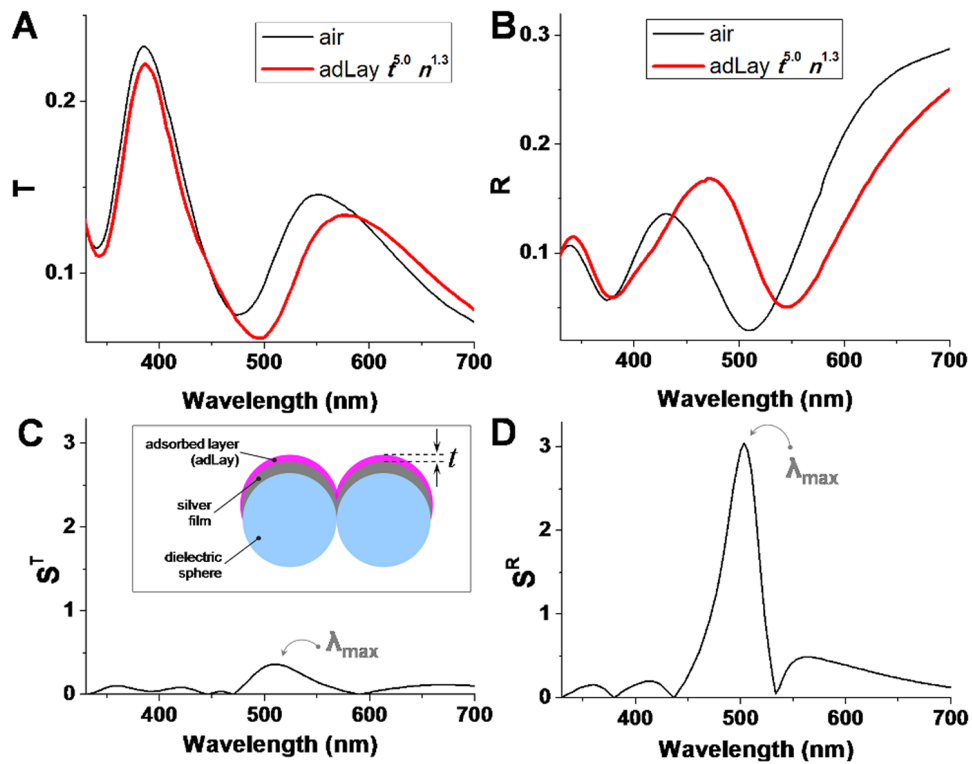

**Figure 2.** (**A**) Transmittance and (**B**) reflectance spectra of bare MCM and MCM coated by a dielectric adsorbate layer, of thickness $t = 5$ nm and refractive index $n = 1.3$. (**C**) $FOM_{layer}$ in transmission mode; (**D**) $FOM_{layer}$ in reflection mode. Inset of (**C**) depicts the thin layer adsorbed (adLay) on the silver film over microspheres.

Previous studies on metal-coated colloidal monolayers[22] have shown that these exhibit a main transmission band similar to the so-called extraordinary/enhanced optical transmission (EOT)[19], in which propagative surface plasmons mediate the light transmitted from one side of an optically thick nanostructured metal film to the other side. For MCM, this EOT band is located immediately to the long-wavelength side relative to the transmission dip observed also in the corresponding bare colloidal crystal[14]. This transmission dip is attributed to light diffracted into the plane of the microsphere lattice, and known as Wood anomaly or Rayleigh-Wood anomaly[18]. The EOT band is the result of the photonic resonance guided mode, propagating along the chain of dielectric spheres, that couples with plasmons on the metal film coating, as demonstrated by previous studies[14,22]. In the transmission spectra on Fig. 1B, the transmission band at 550 nm (denoted as $\lambda^T_1$) represents a supplementary band, of lower intensity, well separated from, and situated at longer wavelengths than the main EOT transmission band at 386 nm (denoted as $\lambda^T_0$). This band at $\lambda^T_1$ shall be reffered to as the secondary EOT band. In turn, the reflectance spectrum is dominated by a main minimum at 509 nm (denoted as $\lambda^R_0$), indicating a resonance at which the overall absorption and enhanced fields on the surface of this structure are at maximum[14]. A question could arise concerning the role played by the triangular metal nanoparticles formed on the substrate in this kind of structure. To answer it, we performed a set of simulations in which the metal particles on the substrate were removed, and compared the optical spectra with those of the whole structure. Results indicate a negligible role of the triangles in the overall spectra (see Supplementary Information). This is probably due to the rather broad plasmon resonances, and less intense than in the photonic-plasmonic hybrid structure of metal-coated microspheres. In fact, although the optical properties of metal-coated monolayers were pretty well explored in the past, this was never done for spheres as small as the ones in this study. Since the optical response of these structures scale with the diameter, the choice of smaller spheres allowed us to obtain more features in the visible range, and to identify some new features, such as the long-wavelength, secondary EOT band, $\lambda^T_1$ (Fig. 1B). We have identified in the literature some transmission spectra that apparently look similar[23], but were obtained on structures that are different than the ones reported here: the spheres were larger and assembled as multilayers, structures also known as metal-film terminated opals[24]. As such, previous works did not identify or analize the secondary EOT band, and did not exploit its sensing potential. As we will demonstrate in the following section, this secondary transmission band is more efficient for sensing than the previously-known main EOT band.

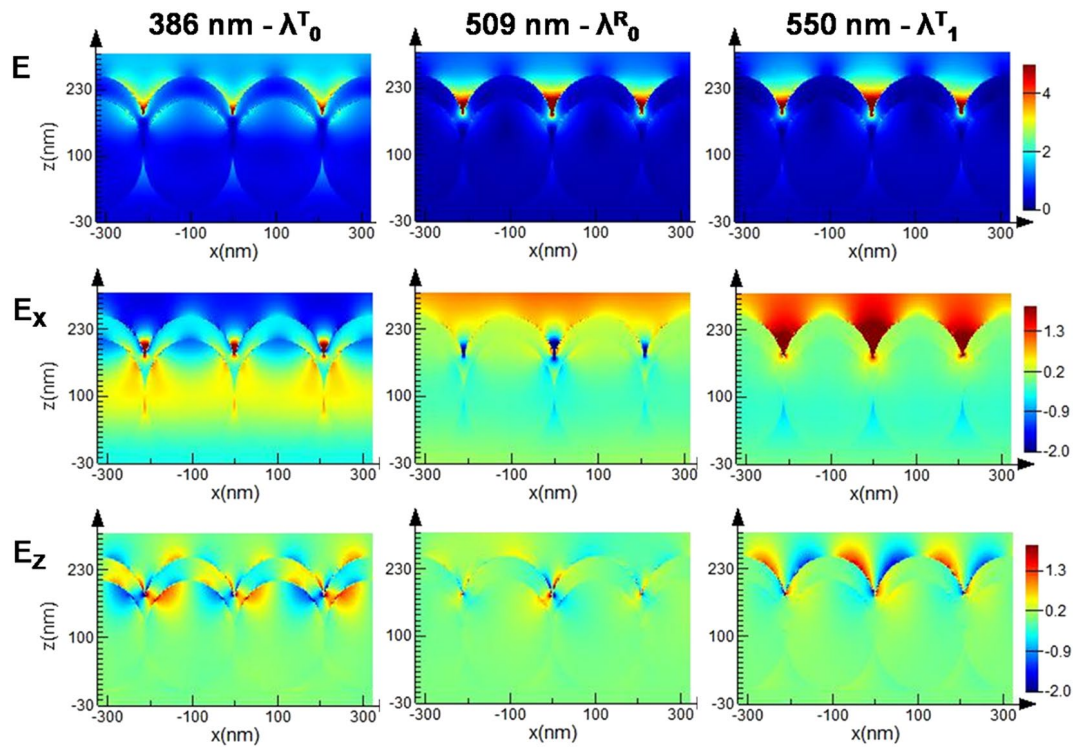

**Figure 3.** Maps of the electric field magnitude $E$ (top row), $E_X$ (center row), and $E_Z$ (bottom row) at the wavelengths of main transmittance maximum $\lambda^T_0$ (left column), main reflectance minimum $\lambda^R_0$ (center column) and secondary transmittance maximum $\lambda^T_1$ (right column).

**Transmission vs Reflection sensing configuration.** Generally, plasmonic sensors rely on the shift of the plasmon resonance position $\lambda_{res}$ induced by small changes $\delta n$ of the refractive index of the medium in contact with the sensor's surface. The resonance can be indicated by either a peak (local maximum) or a dip (local minimum), depending on the properties of the plasmonic nanostructure, and is monitored by a specific imposed setup, configured for transmission or reflection analysis[11]. The sensitivity $S$ is often expressed as $S = \delta \lambda_{res}/\delta n$ and measured in nm/RIU[25]. However, different kinds of nanostructures have broader or narrower plasmon resonance bands $\Gamma$ (full width at half maximum), with shifts of narrower resonances being easier to detect. Thus, a dimensionless quantity was introduced in order to characterize a sensor's quality, namely the 'figure of merit' defined as $FOM = S/\Gamma$. Depending on morphology of the nanostructures, the optical spectra can become quite complex (e.g. non-Lorentzian line shapes), which hampers a straightforward determination of the FOM. However, from a practical standpoint, a spectral shift of the resonance can be detected as a relative intensity change $\delta I/I$ at a fixed wavelength $\lambda$.

Figure 2A,B presents the calculated transmittance/reflectance spectra for the bare MCM and MCM coated by a thin dielectric layer (index 1.3, thickness 5 nm). As it can be observed, the thin dielectric layer induces strong spectral changes both in transmission and reflection spectra. Notably, for the transmission configuration, the spectral region near the secondary transmittance maximum $\lambda^T_1$ is much more sensitive than the main transmission maximum $\lambda^T_0$.

In the reflection configuration, the main reflection minimum $\lambda^R_0$ marks the spectral range most sensitive to changes of the nanoscale vicinity. In Fig. 2C,D the relative variation of transmittance/reflectance is plotted. The remarkable result here, is that the reflection mode can be at least 8 times more efficient for sensing than the transmission mode if a judicious selection of the working wavelength is made. Also, this wavelength $\lambda_{max}$, at which $\delta I/I$ is at maximum is located in the same spectral region for the two configurations (503 nm for reflection and 510 nm for transmission).

Further, in order to better understand and underpin the sensing characteristics of MCM, the electric field distributions are analyzed at selected relevant wavelengths. Figure 3 presents the electric field magnitude $E$, together with the $E_X$ and $E_Z$ components in a cross section along the $XZ$ plane (marked in Fig. 1A) at the wavelengths of main transmittance maximum $\lambda^T_0$, main reflectance minimum $\lambda^R_0$, and secondary transmittance maximum $\lambda^T_1$.

First, the electric field intensity on the top metallic surface (where the adsorbate to be sensed comes into contact) is visibly larger at $\lambda^R_0$ and $\lambda^T_1$ than at $\lambda^T_0$. Secondly, the electric fields at $\lambda^T_0$ are located more towards the inside of the plasmonic structure, on the lower surface of the metal coating and inside the dielectric spheres (see e.g. $E_Z$). On the contrary at $\lambda^R_0$ and $\lambda^T_1$ the fields extend more above the metal surface. Finally, the $E_Z$ field distribution at $\lambda^T_1$ is similar to the one at $\lambda^T_0$, with periodic lobes along the Z direction, indicating the presence of the propagating photonic-plasmonic mode involved in the EOT. However, the location of the fields suggests that, while at $\lambda^T_0$ excitation of plasmon polaritons at the polystyrene-Ag interface takes place, at $\lambda^T_1$ plasmons excited at the Ag-air interface are involved. Thus, the secondary EOT band arises also through excitation of plasmon polaritons in the periodic array. Since at $\lambda^T_0$ the coupling with the photonic guided mode of the sphere array is much stronger, the efficiency of the EOT process is higher, thus the higher intensity of the main transmission band compared to the secondary EOT band. On the contrary, at $\lambda^R_0$, fields are strongly localized at the sharp V-shaped tapered groove formed by the junction between adjacent spheres. All these observations can explain why the main $R$ minimum and the secondary $T$ maximum are efficient for sensing: their associated field distribution extends into the spatial regions in which the adsorbates come in contact with the sensing area. Moreover, the nature of the different modes, i.e. localized plasmons at $\lambda^R_0$, and propagative hybrid photonic-plasmonic modes at $\lambda^T_0$ and $\lambda^T_1$, are in favour of the observed behaviour. It is known that localized plasmons support higher field enhancements than the propagative ones.

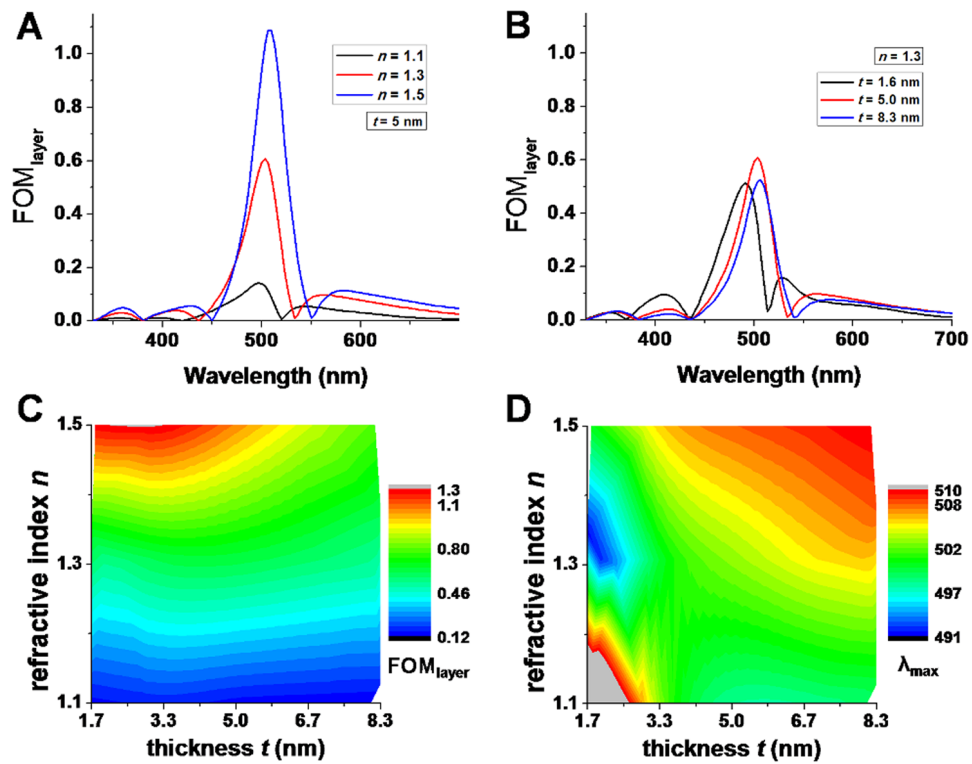

**Figure 4.** (**A**) Dependence of $FOM_{layer} = (\delta R/R_0)/t$ on refractive index $n$ of the adsorbate layer, for thickness $t = 5$ nm. (**B**) Dependence of $FOM_{layer}$ on the thickness $t$ of the adsorbate layer, for $n = 1.3$. (**C**) $FOM_{layer}$ maximum value as function of $n$ and $t$. (**D**) $\lambda_{max}$ value as function of $n$ and $t$. All results are for reflection configuration.

### Dependence of sensitivity on analyte layer thickness and index.

In real-life biosensing applications, the spectrum intensity change is induced by an index change $\delta n$ spatially restricted to a thin layer of thickness $t$ at the metallic surface. Therefore, inspired by the work of Becker et al.[26], we further analyze the sensitivity of the MCM plasmonic sensor in terms of the figure of merit for thin layers, defined as $FOM_{layer} = (\delta I/I)/t$, where $I$ can be either a $T$ or $R$ value. This alternative figure of merit was proposed in order to account for the fact that induced spectral modifications depend on the size of the molecules relative to the volume of the plasmonic modes, and the decrease of sensitivity with increasing distance from the metal surface. Next, $FOM_{layer}$ is evaluated for different values of refractive index $n$ (from 1.1 to 1.5) and thickness $t$ (from 1.6 nm to 8.3 nm) of the thin layer adsorbed on the metal surface. As demonstrated above, the reflection configuration is much more efficient than the transmission one for sensing, thus more detailed analyses are further presented for the former.

The dependence of $FOM_{layer}$ on refractive index and thickness of the adsorbed layer is presented in Fig. 4A and B, respectively. The maximum value of $FOM_{layer}$ increases strongly with the increase of $n$ when $t$ is fixed at 5 nm (Fig. 4A), while for $n$ fixed at 1.3 smaller variations are observed upon increasing $t$ (Fig. 4B). The $FOM_{layer}$ maximum value remains in a relatively narrow spectral region around 500 nm. To better comprehend the dependence of $FOM_{layer}$ on $n$ and $t$, a 2D color-coded map is presented in Fig. 4C. It can be observed that the MCM exhibits a maximum sensitivity for thin layers (3–5 nm) and high refractive index, and that for higher refractive index, the dependence on thickness is stronger.

Interestingly, $\lambda_{max}$, the wavelength of maximum $FOM_{layer}$, slightly shifts towards higher values (490 nm to 510 nm) with the increase of $n$ or $t$ across the explored $n$-$t$ parameter space, as seen in Fig. 4D. This has a huge practical importance, since $\lambda_{max}$ is the wavelength at which measurements should be made in order to optimize a given experimental setup.

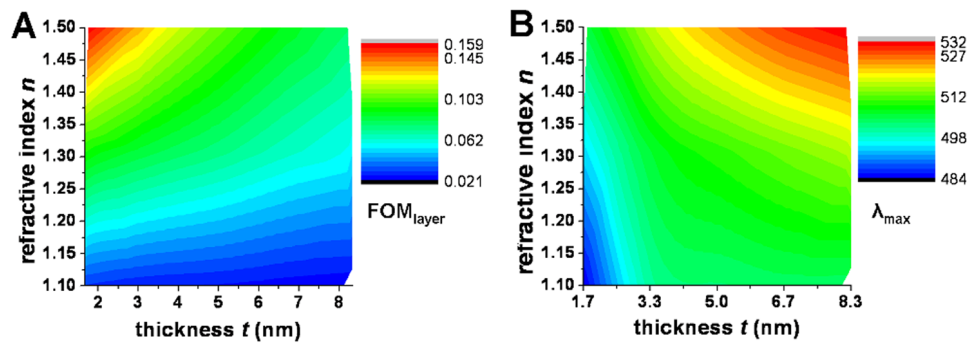

**Figure 5.** (**A**) $FOM_{layer}$ maximum value as function of $n$ and $t$. (**B**) $\lambda_{max}$ value as function of $n$ and $t$. Results are for the transmission configuration.

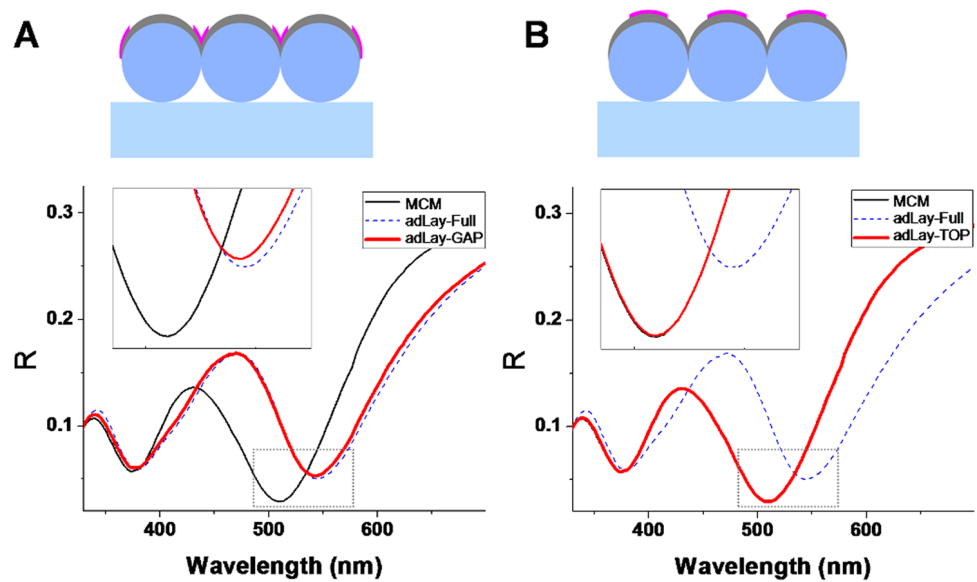

**Figure 6.** Reflectance spectra of MCM partially coated by a dielectric layer located at the gaps (adLay-GAP) between the metal half-shells (**A**) or on their tops (adLay-TOP) (**B**), compared to the spectra of bare MCM (MCM) and fully-coated MCM (adLay-Full).

Figure 5 presents the dependence of $FOM_{layer}/\lambda_{max}$ on refractive index and thickness of the adsorbed layer, for the transmission configuration. The observed trends are similar to those observed for the reflection configuration, but with lower $FOM_{layer}$ values. The maximum value of $FOM_{layer}$ increases strongly with the increase of $n$ and it decreases with the increase of $t$. It can be observed that $\lambda_{max}$, the wavelength of maximum $FOM_{layer}$, shifts towards higher values (from 484 nm to 532 nm) with the increase of $n$ or $t$ across the explored $n$-$t$ parameter space. This shift of $\lambda_{max}$ is higher than for the reflection configuration.

**Dependence of sensitivity on analyte location.** An issue that can often be overlooked in the design of plasmonic nano-biosensors is the fact that beyond having a high sensitivity, the sensor's surface should be easily accessed by the analyte. Highly sensitive plasmonic nanostructures might have complex surface topography, which physically hampers the access of the analyte molecules or bio-nano-objects (e.g. proteins) to the entire area of the sensing surface. The important concern arising from this consideration is whether all the sensor surface area is equally useful for sensing. If the analyte adsorbs/binds to certain or other nanoscopic areas, will the sensor exhibit the same response? In order to investigate this issue a set of simulations were performed with an incomplete adsorbed thin layer: in one scenario the analyte is located only along the sphere equator region, inside the V-shaped grooves (where adjacent spheres and silver half-shells meet) and near the half-shells edges. In the second case the analyte is located only on the top, around the vertical central axes of the silver half-shells.

Results are displayed in Fig. 6 and indicate that the V-shaped grooves formed between adjacent metallic half-shells, and edges around them are the regions sensitive to analyte adsorption. Meanwhile the top of the metal surface is mostly insensitive. In fact this result can be well correlated with the electric field distributions presented in Fig. 3 and discussed above: the analyte needs to reach the regions of enhanced electromagnetic fields in order to be efficiently detected. Obviously, in a typical experiment, all the metallic surface is exposed and available for analyte adsorbtion/binding; however it is crucial that the more sensitive regions are easily reachable by the analyte. The spatial selectivity observed here can have profound implications for practical situations in which large bio-nano-objects are to be detected. Such objects can sometimes be rather rigid so they cannot penetrate into the useful sensing region, thus their presence will be detected with very low efficiency. In situations like this, a careful analysis of the impact of analyte location on the sensor sensitivity, taking into account the nanoscopic morphology of the sensing area, should be conducted. By adjustments of the metallic surface topography, this could lead to a sensor with an optimized morphology, which allows, or even directs the analyte towards the specific locations of interest. For metal-coated microspheres for example, by an additional fabrication step, consisting in etching of the polymer microspheres before the metal deposition, very small gaps can be produced[27,28]. These will be open nano-spaces, and could be easier accessed by a large analyte than the V-shaped grooves found between two adjacent non-etched spheres in the standard metal-coated microspheres.

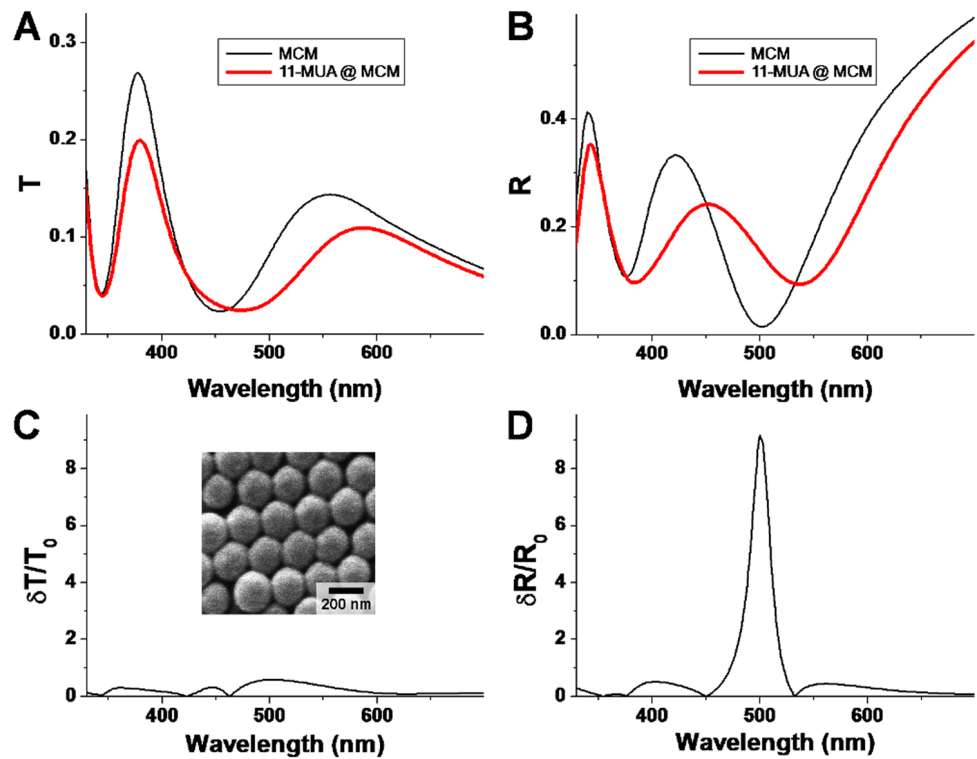

**Figure 7.** Experimental transmittance (**A**) and reflectance (**B**) spectra of bare MCM (black/thin line) and MCM coated by a molecular layer of 11-MUA (red/thick line). (**C**) Relative transmittance variation $\delta T/T_0$. (**D**) Relative reflectance variation $\delta R/R_0$. The inset in (**C**) is a SEM image of MCM.

**Experimental validation: sensing of molecular monolayers.** In order to provide a validation of the results obtained by simulations, sensing experiments were performed with MCM onto which 11-MUA molecular layers were chemisorbed. Beside constituting a model molecule due to its ability to self-assemble onto noble metal surfaces, 11-MUA is also relevant for biosensing as it is a type of molecule often used for cross-linking of the carboxyl to amine groups by the so-called EDC-NHS protocol[29]. Figures 7A and B present the transmittance and reflectance spectra of the fabricated MCM before and after coating the silver surface with a layer of 11-MUA. The spectral shifts induced by the molecular layer follow the trends observed theoretically: both the main $R$ minimum and the secondary $T$ maximum shift to the red ($\lambda^R_0$ from 502 to 538 nm, and $\lambda^T_1$ from 556 to 587 nm), while the main $T$ maximum $\lambda^T_0$ barely shifts about 2 nm (from 378 to 380 nm).

Concerning the observed peak shifts, it is worth noticing that the 2 nm shift of the main transmission band is comparable to a 4 nm shift reported on gold-coated larger spheres (419 nm) onto which 1-dodecanethiol molecular monolayer was adsorbed[21]. Therefore, it can be stressed once more that the secondary enhanced transmission band observed and discussed here, with the peak shifting by 21 nm is more efficient for sensing in the transmission configuration. The relative variation of intensity for the transmission and reflection configurations is also displayed in Figures 7C and D. Again, a very good agreement is obtained with the theoretical findings presented in Fig. 2C,D: similar dependence on wavelength, and a much higher sensitivity of the reflection configuration. The wavelength of maximum sensitivity, $\lambda_{max}$, is here 500 nm, matching those found theoretically. As example, by assuming that the 11-MUA molecular layer has an effective refractive index of 1.5, and analyzing the map in Fig. 4D, one finds that for a $\lambda_{max}$ at 500 nm, the thickness of the film would be in the 1–2 nm range, which is the correct range for 11-MUA molecules. In practice, having an experimental calibration for a given type of sensor with respect to the index $n$ and thickness $t$ of adsorbed layers (like in Fig. 4C,D), could serve as a basis for precisely determining the thickness or index of an unknown layer. The SPR sensing capabilities evidenced above could be further expanded, for example in multimodal sensing approaches, in which SPR sensing could be combined with SERS or electrochemical sensing (the metal film coating the microspheres is conductive) on the same MCM platform. Another possible concern can be the choice of silver in this study, which is prone to oxidation, sometimes hampering the practical implementation of silver based sensors. However silver surfaces can be used fresh, or coated by thin layers of gold. MCM built with silver for the metal coating were proposed here because silver has a wider range of usable optical response across the optical spectral domain than Au. However, an obvious alternative would be gold-coated microsphere monolayers. To enquire about the validity of the obtained results in the case of gold, a set of simulations on MCMs made with gold were performed. Results prove that gold-coated microspheres can also function in both transmission and reflection, and that reflection is more sensitive, as is the case for Ag (see Supplementary Information). Thus the main findings in our manuscript are expected to be generally valid, independent of the noble metal of choice, making their practical implementation highly feasible.

## Conclusions

Metal-coated microsphere monolayers based on small microspheres (210 nm) exhibit multiple spectral features across the visible frequencies region. The sensitivity of both transmission maxima and reflectivity minima to changes of the nearby medium makes them promising for sensing in both configurations. A secondary, long-wavelength enhanced transmission band was identified and its associated electric field distributions analyzed. This newly observed transmission band is more efficient for sensing than the previously investigated main EOT band. The reflection configuration is demonstrated to be about one order of magnitude more sensitive than the transmission one. A figure of merit for thin layers, $FOM_{layer}$ was determined for both configurations, for layers of different refractive index and thickness. Higher index films are generally more easy to detect, while the region 3–5 nm provides the highest $FOM_{layer}$ values, especially at higher index. It is also highlighted that the nanoscopic location at which the analyte adsorbs on the plasmonic sensor's surface plays a crucial role in the sensor's capability to detect that analyte. These new insights contribute to the fundamental understanding of the sensing properties of metal-coated microsphere monolayers. More generally, they can provide guidelines for the rational development of sensitive surface plasmon resonance (bio)sensors based on inexpensive self-assembled colloidal platforms.

## Methods

**FDTD simulations.** Finite-difference time-domain simulations (FDTD) are performed by Lumerical's FDTD Solutions commercial software. Full 3D, realistic structures are modeled: a glass substrate of refractive index 1.52, dielectric spheres (refractive index 1.59) of diameter $D$ (200 nm–500 nm), a silver coating of thickness 45 nm on top of the spheres, an array of silver triangular shaped nanoparticles on the substrate, having a height of 45 nm. Silver coefficients are taken from the CRC Handbook of Chemistry & Physics, and are included in FDTD Solutions' database. A finite hexagonal array, consisting of 95 spheres, was included in the rectangular simulation volume. To model a surface adsorbed layer, an additional dielectric coating (having thickness $t$ and refractive index $n$) is placed on top of the metal surface. A broad-band (300–700 nm) plane wave source illuminates the structure from the air side, transmitted (reflected) intensity being measured below in the substrate (above in the air, behind the light source). Incident light is polarized along the $X$ direction (see Fig. 1A). In order to match experimental conditions, only transmission/reflection into the zeroth diffraction order was analyzed. Note that, while for transmission this is an accurate choice, it is only an acceptable compromise for the reflection case, which experimentally is performed at quasi-normal incidence.

**Fabrication of metal-coated microsphere monolayers.** Monolayers of polystyrene spheres (210 nm diameter, Polysciences) were prepared from aqueous suspensions (2% vol) by convective self-assembly (CSA) on a microscope glass slide. CSA is applied by a home-made equipment; the procedure and setup were described in previous work[27]. Silver films were deposited by thermal evaporation under vacuum, to a thickness of about $45 \pm 5$ nm, monitored by quartz crystal microbalance.

**Sensing experiments.** A monolayer of molecules was adsorbed on the metal surface by immersion of the MCM samples in a $10^{-5}$ M ethanol solution of 11-mercaptoundecanoic acid (11-MUA, Aldrich) for 6 hours, followed by rinsing in ethanol. Transmittance ($T$) measurements were performed at normal incidence in unpolarized light, on a Jasco V-530 spectrophotometer. Reflectance ($R$) was measured at quasi-normal incidence on the same instrument, equipped with a SLM-468S reflectivity extension. Optical spectra are measured both before and after the molecular adsorbate layer was formed.

### Acknowledgements

This work was supported by the Romanian Ministry of Research and Innovation through the National Core Program No. PN19-35-02-01. Publication costs were supported by the Romanian Ministry of Research and Innovation through Program 1 - Development of the national research and development system, Subprogram 1.2 - Institutional performance - Financing RDI excellence, Project no. 32PFE/18. The author thanks M. Gabor at the Technical University Cluj-Napoca for help with silver film evaporation.

### Author Contributions
C.F. designed and performed the experiments and simulations, analysed results, and wrote the manuscript.

### Additional Information
**Supplementary information** accompanies this paper at https://doi.org/10.1038/s41598-019-40261-x.

**Competing Interests:** The author declares no competing interests.

**Publisher's note:** Springer Nature remains neutral with regard to jurisdictional claims in published maps and institutional affiliations.